\documentstyle[12pt]{article}
\begin{document}
\vskip 0.1in
\centerline{\Large\bf Liouville Field Theory and Dark Matter}
\vskip .7in
\centerline{Dan N. Vollick}
\centerline{Department of Physics and Astronomy}
\centerline{Okanagan University College}
\centerline{2552 Trans Canada Hwy NE}
\centerline{Salmon Arm, B.C.}
\centerline{V1E 4N3}
\vskip .9in
\centerline{\bf\large Abstract}
\vskip 0.5in
One of the most intriguing unsolved problems in modern cosmology involves the
nature of the observed dark matter in the universe. Various hypothetical 
particles have been postulated as the source of the dark matter, but as of yet
none of these particles have been observed. In this paper I present an
alternative view. The observations indicate the presence of a new force
that is much weaker than gravity on small scales and stronger than gravity
on large scales. This new force is described by a Liouville field that
consists of a scalar field with an exponential potential. The solutions
have a Newtonian behavior on small scales and produce forces that are 
weak compared to gravity. On large scales the field has a logarithmic
behavior and will therefore produce flat rotation curves.
The Liouville field may therefore be able to account for the observations
that are usually taken to imply the existence of dark matter.
\newpage
\section*{Introduction}
One of the interesting unsolved problems in modern cosmology concerns the
nature of the dark matter in our universe. Observations of spiral galaxies
show that at large distances from the galactic center
rotation curves level off instead of having a 
Keplerian fall off \cite{Tr1,Ko1,Bi1}. 
If this behavior is attributed to dark matter we must
conclude that about 90 percent of the mass 
of spiral galaxies is dark. Observations 
of the velocity distribution in galactic clusters yields similar results.
More recently, observations of the CMBR \cite{Ja1,La1,Me1}
and of type Ia supernovae \cite{Pe1,Ri1,Pe2} have shown
that $\Omega_{total}\simeq 1$ with $\Omega_{DM}\simeq 0.3$ and $\Omega_{DE}
\simeq 0.7$, where $\Omega_{DM}$ is the density parameter for dark matter
and $\Omega_{DE}$ is the density parameter for dark energy (also known as 
quintessence). For excellent reviews of these observations see
\cite{Tu1,Ba1,Tu2}.
  
Of course, all of the above conclusions assume that Newtonian gravity (or
general relativity) is valid. Another possibility pursued by Milgrom 
\cite{Mil1,Mi2} and
others involves modifying Newtonian dynamics. In this theory $a=GM/r^2$ is 
replaced by
\begin{equation}
\mu(a/a_0)a=GM/r^2
\end{equation}
where $a_0\simeq 10^{-10}$ m/s$^2$, $\mu(x)\simeq 1$ when $x>>1$ and $\mu(x)
\simeq x$ when $x<<1$. This allows one to reproduce the rotation curves of
spiral galaxies and some of the 
other observations attributed to dark matter.
  
In this paper I will consider a third possibility: there is a new field
that produces forces which are much weaker than gravity on small scales
and which are stronger than gravity on large scales. The specific field that I
will propose is a scalar field with an exponential potential, that is a 
Liouville field theory \cite{Li1}. 
I will show that near a point source the field behaves 
as $\phi =\phi_0 +a/r$ and at large distance it behaves as $\phi=-\ln(r/r_0)$,
where $a$, $\phi_0$ and $r_0$ are constants.  The Liouville force is given
 by $\alpha c^2\vec{\nabla}\phi$, so that at large distances the
rotation curves will be flat. At small distances the Liouville force will be 
much weaker than gravity if $|\alpha a|<<GM/c^2$. 
Recently there has been another approach \cite{Ma1,Ma2,Gu1}
to the dark matter problem that uses a scalar field and an exponential or
hyperbolic potential. This approach differs from the one presented here in that
their scalar field does not directly exert forces on particles. This scalar
field is a form of dark matter as it modifies the spacetime through its
energy-momentum tensor and this in turn effects the motion of particles in
the spacetime.

Exponential potentials appear in non critical string theory and in
dimensionally reduced spacetimes with a cosmological constant. It will be
shown in the last section that these theories do not produce Liouville
fields with the appropriate parameters to be viable candidates for the 
theory investigated here.
  
Throughout this paper I will use the signature (-+++).
\newpage  
\section*{Field and Particle Equations of Motion}
Consider a collection of massive particles interacting with a Liouville
field $\phi$. The action will be taken to be
\begin{equation}
\begin{array}{c}
S=-\Sigma_nm_n\int\sqrt{-g_{\mu\nu}U^{\mu}_nU^{\nu}_n}d\tau_n+\frac{1}{2}
\Sigma_n\int\lambda_n(\tau_n)[g_{\mu\nu}U^{\mu}_nU^{\nu}_n+1]d\tau_n\\
    \\
+\alpha\Sigma_nm_n\int\phi[x_n(\tau_n)]d\tau_n
-\frac{1}{2\beta}\int[\nabla^{\mu}\phi\nabla_{\mu}\phi-\Lambda^2e^{2\phi}]
\sqrt{g}d^4x\\
\end{array}
\label{action}
\end{equation}
where $x_n^{\mu}(\tau_n)$ and $U^{\mu}_n(\tau_n)$ are the position and 
velocity of the $n^{th}$ particle, $\tau_n$ is the proper time along its
worldline, $m_n$ is its rest mass, $\lambda_n(\tau_n)$ are Lagrange
multipliers and $\alpha, \beta, \Lambda$ are constants.
Note that $\phi$ and $\alpha$ are dimensionless and $\beta$ has the 
dimensions of m/kg.
The field equations are obtained by varying the action with respect
to $\phi(x)$ and are given by
\begin{equation}
\Box \phi+\Lambda^2 e^{2\phi}=-\frac{\alpha\beta}{\sqrt{g}}\Sigma_nm_n
\int\delta^4(x^{\mu}-x^{\mu}(\tau_n))d\tau_n
\label{fieldeqn}
\end{equation}
where $\Box=\nabla^{\mu}\nabla_{\mu}$. The equations of motion for the 
particles are found by varying the action with respect to $x^{\mu}_n
(\tau_n)$ and $\lambda_n(\tau_n)$. They are given by
\begin{equation}
(m_n+\lambda_n)\left[\frac{dU^{\mu}_n}{d\tau_n}+\Gamma^{\mu}_{\alpha\beta}
U^{\alpha}_nU^{\beta}_n\right]+\frac{d\lambda_n}{d\tau_n}U^{\mu}_n=\alpha
m_n\nabla^{\mu}\phi .
\end{equation}
Contracting with $U^n_{\mu}$ gives
\begin{equation}
\frac{d\lambda_n}{d\tau_n}=-\alpha m_n\frac{d\phi}{d\tau_n} .
\end{equation}
Thus, $\lambda=-\alpha m_n\phi$ and the equations of motion for the particles
are
\begin{equation}
\frac{d}{d\tau_n}\left[(1-\alpha\phi)U^{\mu}_n)\right]+(1-\alpha\phi)\Gamma^{\mu}_{\alpha\beta}U^{\alpha}_nU^{\beta}_n=\alpha\nabla^{\mu}\phi .
\end{equation}
In the Newtonian limit (and assuming that $|\alpha\phi|<<1$) the equations
of motion become
\begin{equation}
\frac{d^2\vec{r}}{dt^2}=-\vec{\nabla}\phi_N+\alpha c^2\vec{\nabla}\phi
\end{equation}
where $\phi_N$ is the Newtonian potential and I have explicitly included
factors of c.
\section*{Solutions of the Liouville Field Equations}
In this section I will examine the spherically symmetric solutions of
\begin{equation}
\nabla^2\phi+\Lambda^2e^{2\phi}=-\alpha\beta M\delta^3(\vec{r}) .
\label{eqn}
\end{equation}
There are two solutions of interest. If $\phi$ is sufficiently negative
the $e^{2\phi}$ term will be negligible and the approximate solution
will be
\begin{equation}
\phi=\phi_0+\frac{a}{r}
\label{newt}
\end{equation}
where $a$ and $\phi_0$ are constants. Integrating equation (\ref{eqn})
over a small spherical volume gives
$a=\alpha
\beta M/4\pi$. For this to be the solution near the origin $\phi<<0$
and this implies that $\alpha\beta <0$. However, if we do not allow
true point particles in the theory the vacuum solution only extends down to
the radius of the object, not to $r\rightarrow 0$. Thus, if $\phi_0$ is
sufficiently negative then $\phi<<0$ outside the object even if $\alpha\beta
>0$. So for $\alpha\beta >0$ a particle will refer to an object whose
size is much smaller than the dimensions involved in the problem under
consideration. Note that for extended objects there will be a correction to
$a=\alpha\beta M/4\pi$ produced by the $e^{2\phi}$ term. For simplicity
I will assume that this correction is small.
   
The second (exact) solution is
\begin{equation}
\phi=-\ln\left(\Lambda r\right) .
\label{ln}
\end{equation}
This solution is source free even though it diverges at the origin.
To see this integrate (\ref{eqn}) over a spherical volume of radius $R$,
\begin{equation}
\lim_{R\rightarrow 0}\int\vec{\nabla}\phi\cdot d\vec{s}+\Lambda^2
\lim_{R\rightarrow 0}\int e^{2\phi}dV=-\alpha\beta M .
\end{equation}
Since the left hand side vanishes it follows that M=0. 
The force on a test particle of mass
$m$ corresponding to this solution is
\begin{equation}
\vec{F}=-\frac{\alpha mc^2}{r}\hat{r} .
\end{equation}
For this to be attractive we require that $\alpha >0$. Thus, if $\alpha 
\beta >0$ we have $\beta >0$ and if $\alpha\beta <0$ we have $\beta <0$.
  
These two solutions, (\ref{newt}) and (\ref{ln}), are important 
because numerical integration of (\ref{eqn}) shows that the solutions
are of the form (\ref{newt}) near the origin and are given by (\ref{ln})
at large $r$. This behavior is easy to show for the logarithmic solution.
Let 
\begin{equation}
\phi=-\ln\left(\Lambda r\right)+\epsilon(r) .
\end{equation}
The perturbation $\epsilon(r)$ satisfies
\begin{equation}
r^2\frac{d^2\epsilon}{dr^2}+2r\frac{d\epsilon}{dr}+2\epsilon=0
\end{equation}
to lowest order in $\epsilon$. The general solution is a linear combination of
\begin{equation}
r^{-1/2}\cos\left(\frac{\sqrt{7}}{2}\ln(r)\right)
\end{equation}
and
\begin{equation}
r^{-1/2}\sin\left(\frac{\sqrt{7}}{2}\ln(r)\right) .
\end{equation}
Thus, if we start near the logarithmic solution and integrate to large $r$ we
will approach the solution. If we integrate to small $r$ we will move away from
the solution.
  
Now consider the observational constraints on the parameters in the Liouville
theory. Since the rotation curves of spiral galaxies are approximately flat for
at large distances the Liouville field must be logarithmic
at these distances
and the force it produces must be stronger than the gravitational force.
Equating the centripetal and Liouville forces gives
\begin{equation}
\alpha=\frac{v^2}{c^2}\simeq 3\times 10^{-7}
\label{alpha} ,
\end{equation}
for $v=150$km/s. 
It is important to note that, for the spherically symmetric
solution, the logarithmic Liouville solution is independent of the 
central mass. This implies that the rotation speed in the flat region
should be the same for all galaxies. Most galaxies do have asymptotic
rotation speeds between 100-300 km/s, but they are not all the same. The
above model is however a little oversimplified. Spiral galaxies are not 
spherically symmetric and this will induce a mass dependence in the 
asymptotic speed. In fact, for the cylindrically symmetric case
(see appendix)
\begin{equation}
v^2=\alpha\left[\frac{\alpha\beta \lambda }{2\pi}-2\right]c^2
\end{equation}
where $\lambda$ is the mass per unit length (or the mass in two dimensions).
This mass dependence will however decrease with distance.
Big bang nucleosynthesis calculations indicate that there is a significant
amount of baryonic dark matter. There is not enough of it to account for
galactic rotation curves ($\Omega_{gal}\sim 0.1$ and $\Omega_{bar}\sim 0.04$),
but it may effect the asymptotic value of the rotation speed. Another possible
explanation is that the potential needs to be modified.
  
In the region where
\begin{equation}
\phi =\phi_0+\frac{\alpha\beta M}{4\pi r}
\end{equation}
the force on a particle of mass m is given by
\begin{equation}
\vec{F}=-\frac{\alpha^2\beta Mc^2}{4\pi r^2}\hat{r} .
\end{equation}
We therefore require that $\alpha^2|\beta|c^2<< 4\pi G$. This gives
\begin{equation}
|\beta|<< 10^{-13}\; m/kg .
\label{beta}
\end{equation}
Below equation (\ref{newt}) I noted that there will be a correction to 
$a=\alpha\beta M/4\pi$ for extended objects. If this correction is not
small for macroscopic objects the above constraint will be changed. This
correction depends on the interior solution and is therefore difficult
to compute. For simplicity I will assume that it is small and adopt the
above constraint.
     
A problem arises if the $\phi$ field has a logarithmic behavior in the 
solar system. To see why this is the case note that the radius at which the
Liouville force equals the gravitational force is given by
\begin{equation}
r=\frac{GM}{\alpha c^2}\sim 4\times 10^9 \; m .
\end{equation}
This is obviously in conflict with observations. Thus, near the edge of
spiral galaxies the Liouville field must exhibit logarithmic behavior
but within our solar system it must not. I will define three regions,
a region in which the field behaves as $\phi\propto 1/r$, a transition
region, and a region in which $\phi\propto\ln(r)$. 
   
The distance to the transition region and to the $\ln(r)$ region depends on
$a$ and $\phi_0$ in the expression $\phi=\phi_0+a/r$. As a simple example 
consider the case $a=0$ and $\phi_0<<0$. It is easy to show that the distance
to the transition region is given by
\begin{equation}
r\sim\frac{1}{\Lambda}e^{-\phi_0} \; .
\end{equation}
Numerical analysis shows that the distance to the
$\ln(r)$ region is given by
\begin{equation}
r\sim\frac{50}{\Lambda}e^{-\phi_0}\; .
\end{equation}
For a point particle $a=\alpha\beta M/4\pi$ and $\phi_0$ is undetermined
without some additional information such as the value of the field at some
point. As discussed earlier there will be corrections to
$a= \alpha\beta M/4\pi$ for extended objects.
However, just as in the point particle case the 
constants $a$ and $\phi_0$ cannot, in general, be determined without some
additional information. This occurs
since it is not possible to impose a boundary
condition at infinity. Thus, without some additional information it is
not possible to predict the distance to the transition region in our
solar system and galaxy.
   
\section*{The Energy-Momentum Tensor}
The energy-momentum tensor for the $\phi$ field,
\begin{equation}
T^{\mu\nu}_{\phi}=\frac{2}{\sqrt{g}}\frac{\delta S_{\phi}}{\delta g_{\mu\nu}} ,
\end{equation}
is given by
\begin{equation}
T^{\mu\nu}=\frac{1}{\beta}\left[\nabla^{\mu}\phi\nabla^{\nu}\phi-\frac{1}{2}
g^{\mu\nu}\left(\nabla_{\alpha}\phi\nabla^{\alpha}\phi-\Lambda^2e^{2\phi}
\right)\right] .
\label{em}
\end{equation}
If $\phi<<0$ we have
\begin{equation}
T^{\mu\nu}\simeq\frac{1}{\beta}\left[\nabla^{\mu}\phi\nabla^{\nu}
\phi-\frac{1}{2}
g^{\mu\nu}\nabla_{\alpha}\phi\nabla^{\alpha}\phi\right]
\end{equation}
which is the energy-momentum tensor for a Klein Gordon field if $\beta>0$.
If $\beta<0$ the field violates the weak energy condition. For $\phi=
\phi_0+\alpha\beta M/4\pi r$ the energy density is ($g_{\mu\nu}=\eta_{\mu\nu}$)
\begin{equation}
\rho=\frac{\alpha^2\beta M^2}{32\pi^2 r^4}\; ,
\end{equation}
which is much smaller than the Newtonian gravitational energy density.
  
In the region where $\phi=-\ln(\Lambda r)$ it is easy to see that
$\nabla_{\alpha}\phi\nabla^{\alpha}\phi-\Lambda^2e^{2\phi}=0$ and
\begin{equation}
T_{\mu\nu}=\frac{1}{\beta}\nabla_{\mu}\phi\nabla_{\nu}\phi .
\end{equation}
Thus, $\rho=0$ and $P_r=T_{rr}$ is given by
\begin{equation}
P_r=\frac{1}{\beta r^2} .
\end{equation}
The spacetime will not be asymptotically flat due to the slow fall off
rate of the pressure.
   
It is interesting to note that violations of the 
weak energy condition can actually occur even if $\beta >0$. 
To see this note that the total energy-momentum tensor is given by \cite{Vo1}
\begin{equation}
T^{\mu\nu}=\Sigma_n\frac{m_n}{\sqrt{g}}\int(1-\alpha\phi)U^{\mu}_n
U^{\nu}_n
\delta^4(x-x_n(\tau_n))d\tau_n+\frac{1}{\beta}\left[\nabla^{\mu}\phi\nabla^
{\nu}\phi-\frac{1}{2}g^{\mu\nu}\left(\nabla_{\alpha}\phi\nabla^{\alpha}\phi
-\Lambda^2 e^{2\phi}\right)\right] .
\end{equation}
For the solution $\phi=-\ln(\Lambda r)$ 
\begin{equation}
T^{00}=\Sigma_n\frac{m_n}{\sqrt{g}}\left(1+\alpha\ln(\Lambda r)
\right)\left(U^0_n\right)^2\delta^4(x-x_n(\tau_n))d\tau_n .
\end{equation}
Since $\alpha>0$, $T^{00}$ will violate the weak energy condition if
particles are placed at $\Lambda r<e^{-1/\alpha}$.
    
Throughout this paper I have assumed that the spacetime curvature produced
by $\phi$ is negligible. If this is not the case the Liouville field would act 
as dark matter and these effects would have to be taken into account. Here I
am taking the point of view that the Liouville field is not dark matter, 
but instead directly exerts forces on objects. Since the energy density in
the $\phi=\phi_0+a/r$ region is much smaller than the Newtonian gravitational
energy density I will focus on the $\phi=-\ln(\Lambda r)$ region. In
this region $\rho=0$ and $P_r$ will be the source of the gravitational field.
In the weak field limit
\begin{equation}
\Box h_{\mu\nu}=-16\pi G\left(T_{\mu\nu}-\frac{1}{2}\eta_{\mu\nu}T\right)
\label{weak}
\end{equation}
where $g_{\mu\nu}=\eta_{\mu\nu}+h_{\mu\nu}$ and $T=\eta_{\alpha\beta}
T^{\alpha\beta}$.
Let $R$ be the radius at which the logarithmic behavior begins for the
Liouville field, and let M be the mass of the galaxy. The Liouville field
will have a negligible contribution inside $R$ (I will neglect the transition
region) and the visible matter in the galaxy will have a small contribution 
outside R. Solving (\ref{weak}) with $h_{00}=2GM/{c^2r}$ just below $r=R$ gives
\begin{equation}
h_{00}=\frac{2GM}{c^2r}-\frac{8\pi G}{\beta c^2}\ln\left(\frac{r}{R}\right)
+\frac{8\pi G}{\beta c^2}\left(1-\frac{R}{r}\right)
\end{equation}
Thus, the gravitational force on a particle of mass $m$ at $r>R$ is
\begin{equation}
\vec{F}\simeq-\frac{4\pi Gm}{\beta r}\hat{r}
\end{equation}
and for the logarithmic Liouville force to dominate we require that
\begin{equation}
|\beta|>>\frac{4\pi G}{\alpha c^2}\simeq 
10^{-20} \; m/kg .
\end{equation}
This inequality can be relaxed if we allow the Liouville field to act
as a source of dark matter. Note that $|h_{00}|=1$ 
at $r\simeq R\exp{[|\beta|
c^2/8\pi G]}$, so that the weak field limit is valid out to large distances.
  
Thus, the final constraints on $\alpha$ and $\beta$ are 
\begin{equation}
\alpha\simeq 3\times 10^{-7}
\end{equation}
and
\begin{equation}
10^{-20}<<|\beta|<< 10^{-13} m/kg
\label{const}
\end{equation}
with the lower limit on $\beta$ only being necessary if the Liouville field
is not to act as a source of dark matter. The upper limit is based ont he 
assumption that $a=\alpha\beta M/4\pi$, which is valid for a point particle.
If the corrections to $a$ are significant for macroscopic objects this limit
will change.
  
Liouville potentials occur in non critical string theory non critical
string theory. In the Einstein frame the effective action for the dilaton
in string theory is \cite{Po1}
\begin{equation}
S_{dil}=-\frac{2}{\kappa^2(D-2)}\int d^Dx\sqrt{g}\left[\nabla_{\mu}\phi
\nabla^{\mu}\phi-\frac{(D-2)(26-D)}{6\alpha^{'}}e^{4\phi/(D-2)}\right],
\end{equation}
where $\kappa^2=8\pi G$ and $D$ is the dimension of the spacetime. The action
is of the form (\ref{action}) for $2<D<26$ with
\begin{equation}
\beta=\frac{\kappa^2}{(D-2)}\; ,
\end{equation}
and
\begin{equation}
\Lambda^2=\frac{2(26-D)}{3\alpha^{'}(D-2)}\; ,
\end{equation}
It is eay see that $\beta$ does not satisfy (\ref{const}), so that the dilaton 
would be a form of dark matter. Secondly, 
$(\alpha^{'})^{1/2}$ is of order of the Planck length, whereas $\Lambda^{-1}$
is expected to be much larger. 
Thus, the string dilaton is not a likely candidate for the
Liouville field discussed in this paper.
   
Liouville potentials can also occur in dimensionally reduced spacetimes
\cite{Mi1}. In this scenario $\Lambda^2$ is related to the cosmological
constant so that it could be astronomical in scale. However, $\beta$ violates
the lower inequality in (\ref{const}) so that it would act as a source
of dark matter as well as exerting a direct force on objects.
\section*{Conclusion}
In this paper I examined an alternative explanation to the observations
that are usually taken to imply the existence of dark matter. In this
approach there is a new force produced by a Liouville field. On small
scales the force is weaker than gravity but on large scales it is stronger
than gravity and produces flat rotation curves. This Liouville field may
therefore eliminate the need for a significant amount of dark matter in
the universe.

\section*{Appendix}
In this appendix I will briefly examine some of the relevant aspects of
the two dimensional circularly symmetric (or
cylindrically symmetric in three dimensions) solutions to Liouville's equation.
The advantage of working in two dimensions is that exact solutions are
known.
A two parameter solution to
\begin{equation}
\nabla^2\phi+\Lambda^2e^{2\phi}=0
\label{eqn2d}
\end{equation}
is given by \cite{Wa1}
\begin{equation}
\phi=-\ln\left[\frac{\Lambda }{2n}r\left[\left(\frac{r}{a}\right)
^n+\left(\frac{a}{r}\right)^n\right]\right]
\end{equation}
where $a$ and $n>0$ are arbitrary constants. Including a point source
(replace the right hand side of (\ref{eqn2d}) by $-\alpha\beta M\delta^2
(\vec{r})$) fixes $n$ to be
\begin{equation}
n=1-\frac{\alpha\beta M}{2\pi} \; ,
\end{equation}
if $\alpha\beta M/2\pi < 1$.
The constant $a$ is left undetermined. In three dimensions the mass $M$ 
becomes the mass per unit length.
  
For $r<<a$ the acceleration is given by
\begin{equation}
\vec{a}\simeq\frac{\alpha(n-1)c^2}{r}\simeq -\frac{\alpha^2\beta Mc^2}
{2\pi r}\hat{r}
\end{equation}
and for $r>>a$
\begin{equation}
\vec{a}\simeq\alpha c^2\left[\frac{\alpha\beta M}
{2\pi}-2\right]\frac{\hat{r}}{r}\; .
\end{equation}
Note that for sufficiently large or small $r$ the acceleration is
independent of $a$. However, for $r\sim a$ the acceleration will depend on $a$.
For gravity to dominate at small $r$ we must have $\alpha^2|\beta|c^2<<2\pi G$.
   
Finally, I want to consider how a circular shell effects the behavior
of a system located its center. 
If a source of mass $M$ at the origin of the shell
the interior solution is
\begin{equation}
\phi_{int}=-\ln\left[\frac{\Lambda}{2n}r\left[\left(\frac{r}{a_1}
\right)^n+\left(\frac{a_1}{r}\right)^n\right]\right]
\end{equation}
with
\begin{equation}
n=1-\frac{\alpha\beta M}{2\pi} .
\end{equation}
Note that this is the same expression for n that one would obtain without
the shell. Thus, the shell can only change the
value of the constant $a$. Since the force is independent of $a$ for
small enough $r$, the shell will not effect the motion of objects
close to the source.

\end{document}